# Enhancing Fine Motor Skills of Wards with Special Needs Using Cluster Model of Cognition

T.R.Gopalakrishnan Nair[1], N. Sowjanya Rao[2], Ananda Bukkambudhi[3]
[1] Director –Research and Industry, Senior Member IEEE,
trgnair@ieee.org
[2] Project Associate, Research and Industry Incubation Center,
sowjan@gmail.com
[1,2] Dayanand Sagar Institutions, Bangalore
[3] Department of Mechanical Engineering, SBMJCE, Bangalore
ananda.bukkambudhi@gmail.com

## ABSTRACT

Technology offers great potential to overcome physical barriers of human race. This paper presents the methods of enhanced learning applicable to children having special needs using better human-computer interaction. The Audio-Visual (AV) effects that the graphic tools or animations help in achieving better learning, understanding, remembering and performance from such students. The 3L-R Cluster Program Model enable them to look into pictures and animated objects while listening to the related audio. It also motivates them to do the FMS development activities like drawing, coloring, tracing etc., certain types of games in the clustered model will help the children to improve concentration, thinking, reasoning, cognitive skills and the eye-to hand co-ordination. Here we introduced a novel cluster model along with the methodology described which provides an ample exposure to the effectiveness of the training. Classify the students with similar problems or disability and the associated curriculum of modified teaching methodology to meet their special needs is met through the specialized IT tool which form a part of the cluster model. It ensures effective learning in wards by enhancing multifaceted interaction. The main objective of this paper is to support the development of Fine Motor Skills (FMS) of wards with special needs.

## INTRODUCTION

Technology could improve access to learning for people with disabilities. Additionally, students with LD (Learning Disabilities) often experience greater success when they are made aware of their abilities (strengths) to work around their disabilities (challenges) [1].

The following quotation from Martin and Kristen [1] is worth noting here.

> "For people without disabilities, technology makes things easier.
> For people with disabilities, technology makes things possible."

Working with IT medium support increases self-determination, independence and integration skills and allows for "positive changes in inter and intrapersonal relationships, sensory abilities, cognitive capabilities, communication skills, motor performance, self-maintenance and leisure productively" [5]. Thus, it is becoming more and more evident that the computer can be used as an effective learning tool to support the acquisition of basic skills.

The literature survey shows that various educational softwares for kids make learning easy for those children who are differently able or with special needs to improve their overall development. Some examples of how computers can be used by children with special needs are listed below [6]:

- *Compensatory Tool:* A PC with speech output can be used to focus the attention.
- *Computer Assisted Instructions:* Tutorial, worksheets and interactive software programs assist the child in better learning.
- *Access to Information:* Computers and networks provide new options for accessing information through on-line books, journals and other such resources.
- *Electronic Communication:* Those with hearing and speech impairments, having a PC with internet can communicate with others using text messages on electronic networks. It can also facilitate social and language development.

However, with the advent of technology, the computer enriched environment is a great boost for everyday life management of people with special needs.

## Motor Skills

It is a common knowledge that children who experience learning disabilities have delayed developments in the motor skills. A motor skill is described as an action which involves the movement of muscles in a body. They are divided largely into two groups:

- Gross Motor Skills
- Fine Motor Skills

Under normal circumstances, both types of motor skills usually develop together because many activities depend on the coordination of both the skills. Gross motor skills, which include the larger movements of arms, legs, feet or the entire body. Gross motor skills enable such functions as crawling, running, jumping walking, kicking, sitting upright, lifting and throwing a ball, maintaining a body balance, coordinating etc., [3] and fine motor skills, which are smaller actions, such as holding an object between the thumb and a finger or using the lips and tongue to pronounce words and syllable. The paper discusses mainly on developing Fine Motor Skills using the model and methodology described with the help of IT tools. Development of Gross motor skills is beyond the scope of this paper.

## Fine Motor Skill (FMS)

Fine Motor Skills can be defined as coordination of small muscle movement which occurs in fingers in coordination with the movement of eyes. The term dexterity is commonly used in application to motor skills of hands and fingers [2]. Simple activities such as writing, drawing, buttoning, faster clothing etc., involve a refined use of the small muscles controlling the hand, fingers and thumb. Motor therapy including speech therapy can be started at an early stage for the development of speech and language. The abilities which involve the use of hands, such as holding an object or catching a ball involve precise hand to eye coordination. Fine motor skills (FMS) can be developed with time and practice.

FMS include [4]

*Ocular Motor control* : The ability of the eyes to follow and focus on an object in the field of vision as required.

*Hand-eye coordination*: The ability to execute activities with hands, guided by the eyes requiring accuracy in placement, direction and spatial awareness.

*Manual dexterity* : The ability to accurately manipulate the hands and fingers for neat handwriting, drawing, typing skills etc.

*Sterognosis* : The ability to recognize unseen object using the sense of touch.

*Tactile perception* : The interpretation of information transmitted via the fingertips to the brain.

## Methodology to develop Fine Motor Skills

The teaching methodology is initially without the aid of computers. Firstly, teaching pre-linguistic skills such as gesture imitation to build a friendly inter-relationship with the child and make the child to listen and obey the instructions given orally. Secondly, teach the child self care skills such as buttoning, zipping/unzipping, carrying etc. to manage day to day life management.

Finally train the child with ease the technology in learning process. Some of the FMS activities like coloring, drawing are taught using computer with IT tools. Developing a skill should be made more interesting by changing slightly each time the position and the activity in a progressive manner.

Before the model based training commences, the percentage of performance of each child in each activity should be given by either the parent or the special educator. This gives the data for the first column before the training in Table 3.

## Teaching FMS without the aid of IT Tools

**i. Pre-linguistic Skills**

This skill starts with use of gesture. Encourage the child to imitate or show with sign what he / she want. This will make the parent to understand the needs of the child. The sooner the training is provided the better is the progress of overall development of the child. The first step involves eye-contact that is making the child to look into the eyes of the person who is talking to him/her. Next, to gain the auditory attention, make small sound or play music and see that the child turns towards the direction from where the music or sound is heard. Let the instructions be simple and short in the beginning to the level of his/her understanding. Make the kid to repeat simple actions including gesture and voice. Each time slightly change the activity to gain the interest and make it appealing and responsive.

**ii. Self-Care Skills**

The self-care skill activities such as buttoning, lacing etc., [2, 3, 4, 11, 12 and 13] are taught which help in day-today life management of the child. These activities improve his/her fine motor skills, self-dependent, develop confidence and make him/her manageable. Each activity procedure should be explained in simple way. If necessary the actions can be shown by the trainer.

- *Buttoning* : Start with bigger buttons and show the kid to button a shirt which is in hanger first. Then guide to button the cloth which he/she has worn.
- *Zipping/unzipping* : Zip or unzip the dress. Make the child to hold the cloth under the zip straight and firm before zipping. Similarly buckling and unbuckling of the dress part or a belt or shoes must be taught.
- *Carrying* : This activity makes the child to feel responsible and caring. Give the child to carry small things like umbrella, vegetables, his/her own water bottle and repeatedly remind to carry it carefully and safely until the destination is reached. Most important is appreciate the child, mentioning that it was of great help to you. Appreciation always builds confidence.
- *Opening and closing* : This activity can be given even in absence of supervision. Let the child try to open or close the plastic box, his/her own food box and water bottle. Let the shape, opening and closing technique is slightly different from the previous ones. Avoid giving glass items or the items which may be broken or have sharp edges and corners. The blocks can also be used in this case.
- *Picking up and holding small objects* : Make the child to practice picking up grains, cereal etc., Edible items should be given in the beginning as a safety measure.
- *Pinching objects between fingers* : This activity builds self confidence. It helps in effective and proper pencil grip.
  Make the child to hold any object between the fingers, encourage and motivate the child by simple actions such as a pat him/her saying that he/she is strong. *Isolate finger*

- *Movements* : This activity can be taught with fun. That is, using one finger at a time such as playing the kid piano or typing or rhymes involving counting.
- *Turning things* : The activity such as turning pages of a book. You should observe that a kid turns only one page at a time and in a proper manner without tearing the paper.
- *Holding and Placing* : Make the child to assist you in keeping utensils/household items properly and holding them effectively. Name each item and ask him/her to bring the item you mentioned. The activity helps in pick and place things properly.
- *Bathing* : Teach the child to bath. Instruct the child to keep all the things ready like towel, soap, paste, brush etc., before he/she starts. This activity helps in organizing and remembering things.
- *Screw and unscrew* : Give this activity under supervision only. It helps in gripping an object and concentration development.
- *Playing games* : The games require precise hand and finger control. Games such as holding a small item in one of the closed hands and asking others or other kids (preferred) to find out/ guess in which hand the item is likely to be found.

  Use play dough to create small balls of near spherical shapes. Make the child to prepare dough for chapatti. Give him/her small amount of flour, little water and ask the child to prepare dough. Same thing can be repeated with clay that is non toxic. However, care must be taken to make the child aware that it is not edible.
- *To improve eye-hand coordination* : The child can be involved in group games such as carom board, dice. Certain games which will help in development of social skills such as friendliness, sharing, loosing, gaining are (i) string beads with bigger holes (who beads the maximum beads in a given time period), (ii) hit the ball to wall and catch it (count the number of catches), (iii) throwing ball or any light weight objects into a tub or a box (specified number of chances are given to each child).

*Clap and Count* : The trainer can clap first and instruct the child to concentrate how many times the clapping was done and ask to repeat clapping for the exact number of times. Develops number concepts and improves the attention span.

- *Tearing paper* : Tearing a newspaper into fine strips and crumple them into balls. Screw up whole piece of a newspaper in one hand gives strength to FMS.
- *Scribbling, Writing* : Scribbling and writing on the board which makes wrist to be bent back gives good exercise to wrist, shoulder and arm muscles.
- *Finger Tracing* : Make the child to trace a pattern in sand or in air, that is in front of his /her eyes with arm outstretched improves the imagination skills, recollecting builds strength in fingers.

## Teaching FMS activities through IT Tools

Computer supported teaching with audio and video media instead of text book based presentation of information will add interest to the learners. Computer assisted teaching is used by the trainer. Multimedia components with highly interactive are the most attractive features of computer supported teaching [9].
Some of the FMS activities like coloring, drawing and tracing can be done with the help of IT tools. The model uses Audio-Visual (AV) tools to serve the purpose. Graphic tools or animation with regard to the skills mentioned above can be used to facilitate learning, understanding, reasoning and performing a task effectively. Each child has to be tested several times depending on his/her concentration attention span with specific software.
The teaching approach, methodology and student performance in cognitive ability have to be regularly recorded. This test is called Cognitive Ability Test (COGAT). The evidence supporting effectiveness can then be shown by means of graphical representation.

## Design Model

The 3L–R cluster program model is designed considering a group of 8-10 kids of special needs with similar disability and approximately same level of difficulties to develop Fine Motor Skills (FMS) such as Ocular Motor control, hand-eye coordination, manual dexterity etc.,

The model is designed on the basis of identifying

- The specific needs and challenges in a group
- The interests, skill and ease in using technology.
- The strengths (abilities) – specific and in coordination

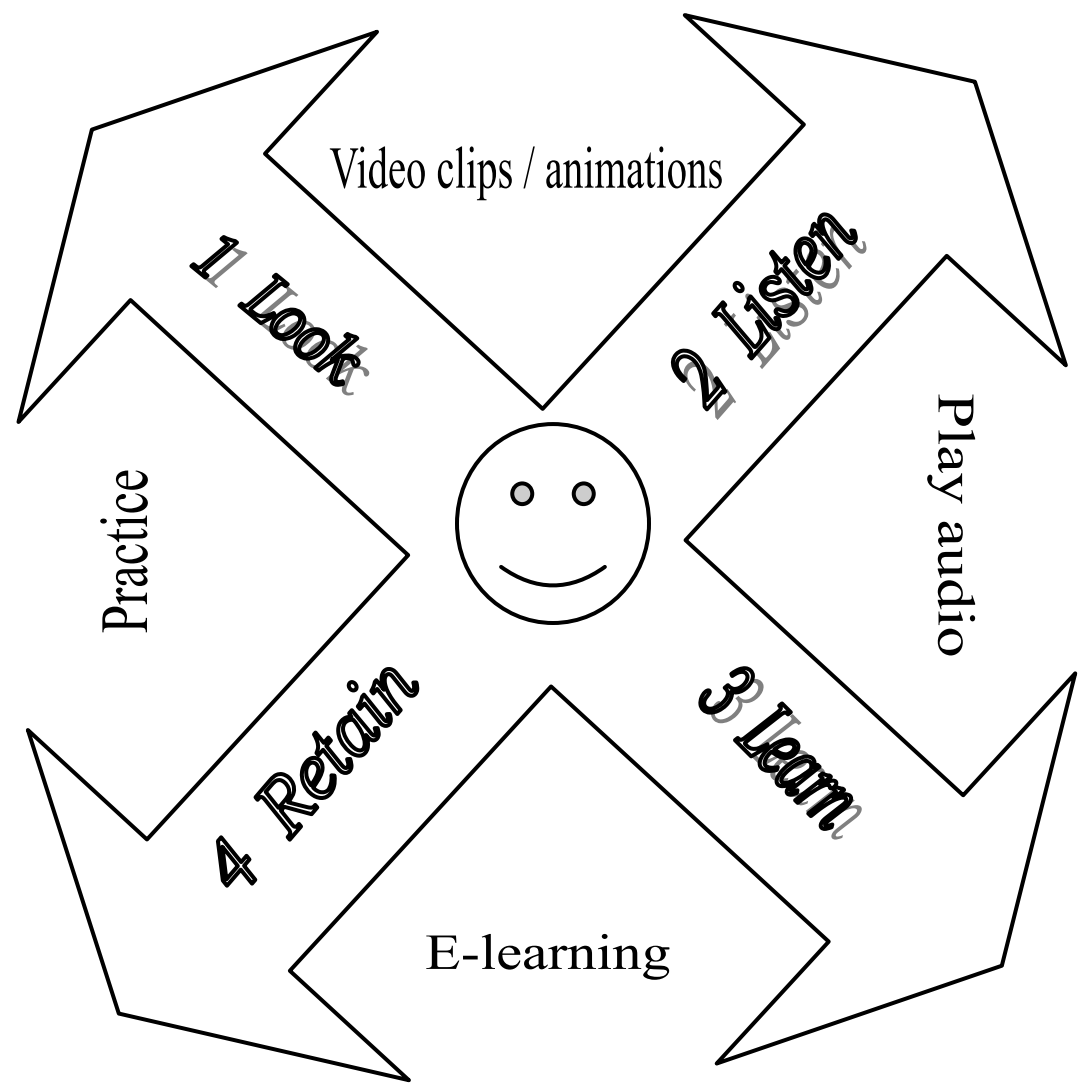

Figure 1. 3L-R Cluster Program Model

The aim of the 3L-R Cluster Program Model is

- Show and tell (Acquisition)
- Practice makes perfect (Fluency)
- You got it – Retain (Maintenance and Generalization) [7]

The 3L-R cluster model is based on AIDCA factors. (Attention → Interest → Desire → Conviction → Action). The mind remembers speech less well than images, and images alone are registered less efficiently than an Audio-Visual (AV) combination [8].

- **Attracting Attention** : Gain attention of the child. Attention can be grabbed by calling their name or play music.
- **Developing Interest** : Turn the attention into interest with the multifaceted activities like showing animated pictures, playing music etc.,
- **Inspiring Desire** : The lesson plans are organized in such a way to make a condition of desire in child.
- **Creating conviction** : Encourage the conviction that the teaching through Audio-Visual (AV) aid is more effective and interesting than the textbook based presentation of information.
- **Inciting Action** : The test of effectiveness is Action. Make the child to Act on the oral instruction given by the trainer.

The focus is on the last 'A' in AIDCA, make the child to Act now.

The model starts teaching with **Look** concept as shown in figure 1 with visual supports and then to **Listen** through related recorded audio. Thirdly, the **E-Learning** with activity sheets, color picture and animation objects support the child to engage in showing interest to the learning. Finally AV support training with practice helps in **Retaining** and applying the skills to new situation, activities and ideas in day to day life management of the child. The model helps in training the children to develop FMS activities such as coloring, pick and place, drawing, games etc., using various softwares, power point slides, worksheets and PDF tutorials.

**Activity 1. Coloring**

This is a simple activity which helps the children learn with fun. It facilitates the child about the color concept, imagination skill and in recognizing the shapes. It develops eye- hand co-ordination precisely. The free educational software like Leay's free coloring book, Mihir, ABC color with me etc., help children to color the ready pictures.

The simplest coloring software opens with ready picture, a brush and set of color icons (color palette) on the screen. The child has to select the color and click the brush inside the picture as shown in figure 2. The picture gets filled up with the selected color. Some of the other software gives pictures with several user-friendly toolbar options to select and draw different picture, shape or object. This activity can be given for a period of 3 to 4 weeks depending on the learning speed and interest of each kid. After the training, give the child, print outs of the pictures on a paper and ask him / her to color it with crayons.

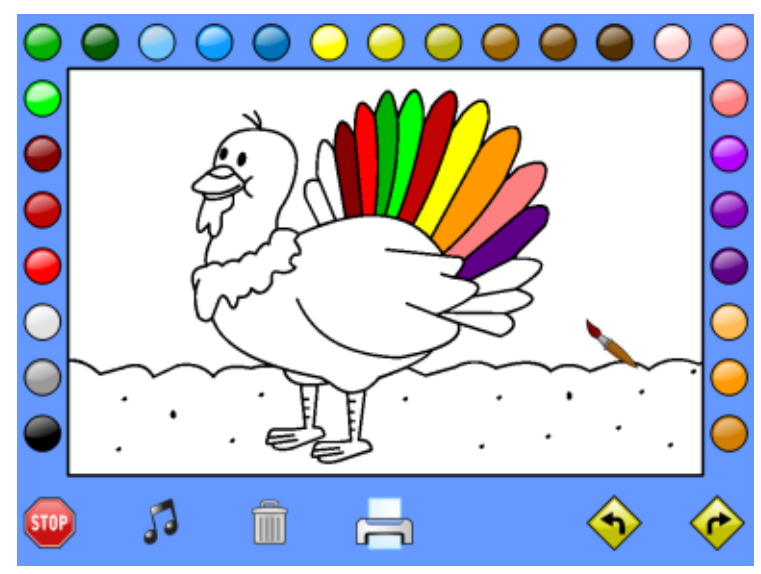

Figure 2. Ready Picture to color

**Activity 2. Placing or Pasting**

This activity improves observation and orderliness. Initially placing the block, paper, cotton, cereal on an outline can be taught manually. Secondly the educational software like MyABCD provides the set of objects in random order as shown in Figure 3. Now the child has to place each shape in its appropriate place to complete the picture of train.

Variety makes a work of art interesting. Make the slides with similar concept using geometrical shapes such as triangle, rectangle or oval. Draw the outline of each shape ten to twenty times in different sizes and different positions. Use same shapes with color on the other side. Ask the child to drag and drop each colored shape in its outline [10]. This activity can be given for a period of two week. The other softwares like Fruit folic and Shape shifter gives the activity to place the object appropriately and precisely.

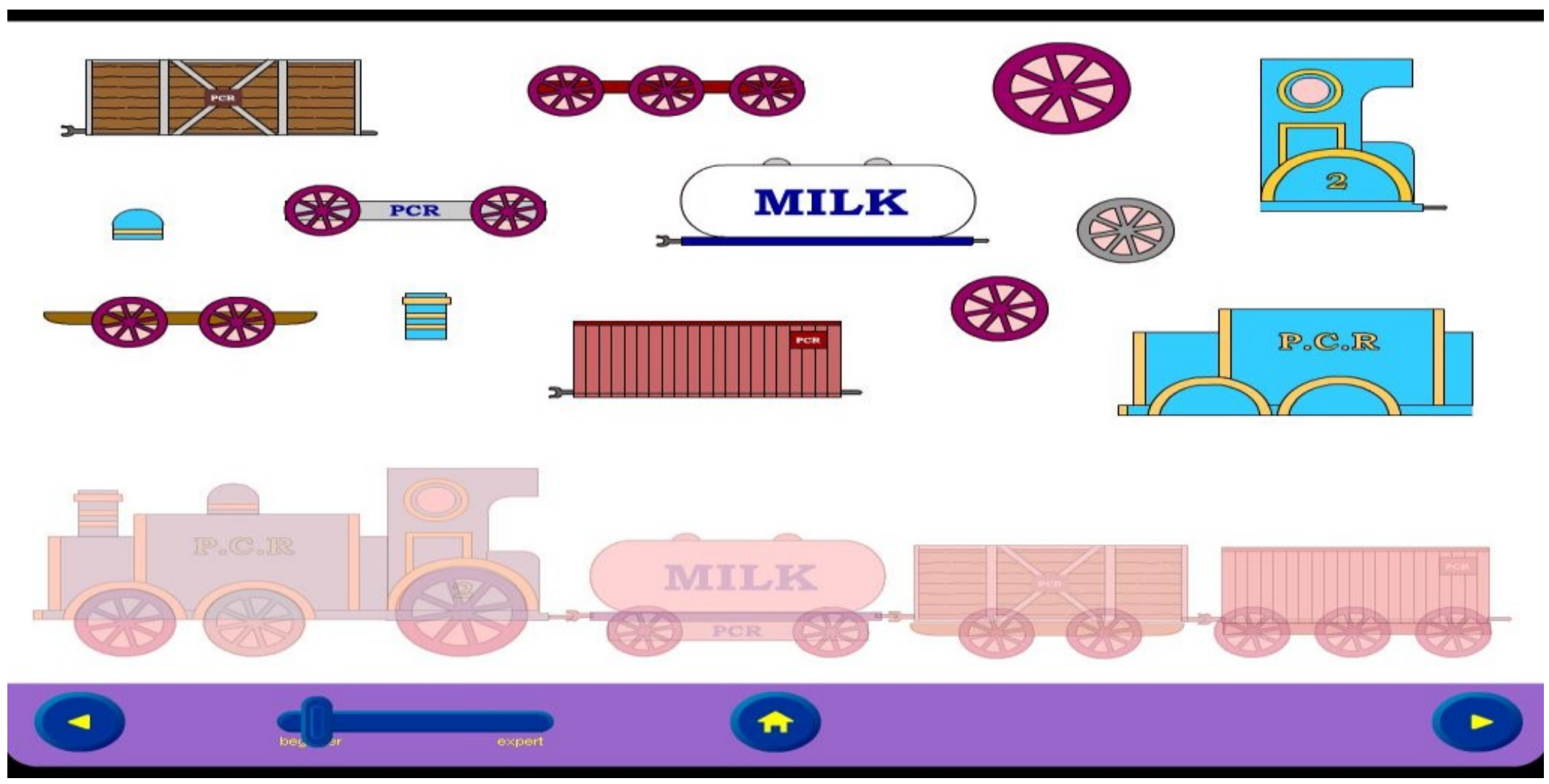

Figure 3. Pick and place the object

**Activity 3. Tracing**

This activity helps in improving writing and drawing activity. Initially basic shapes such as lines, square, circle and alphabets can be given for tracing as shown in figure 4 (a). Use light colored line to guide the tracing. Decrease the size and increase the complexity gradually.

**3.1 Dot Pattern**

Few children find it difficult to differentiate between b and d, p and q, O and Q, M and W etc., which look similar. By repeatedly copying Dot pattern designs, children learn to see the different shapes that make up letters and numbers and so overcome their letter-reversal perception issues (generally found in dyslexia). In order to help students see the letters of the alphabet as a series of lines and curves that form different letters have the students do dot patterns. It could be very useful for students with dyslexia and ADHD [9].

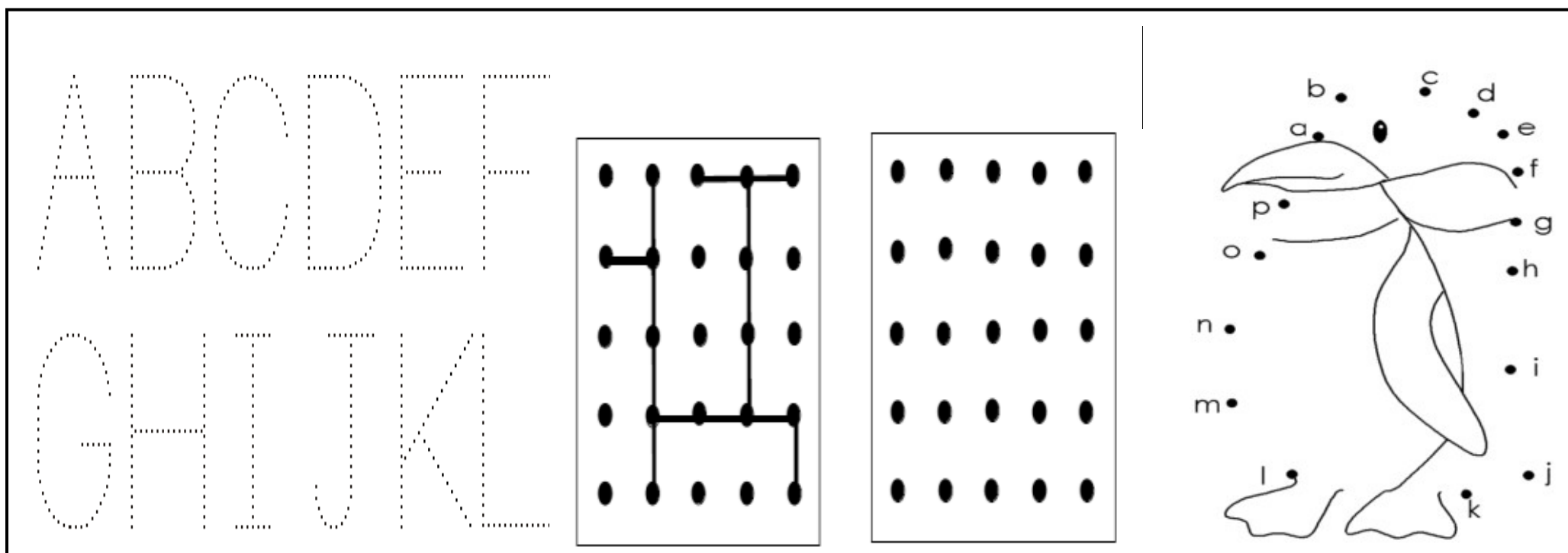

(a) (b) (c)

Figure 4. Tracing and Dot patterns to join the dots

The dot patterns in figure (b) shows the first part as a sample to show the child to join the dots. In the second part of figure 4(b) matrixes of dot is given and instruct the child to join the dots as shown in the first part. Similarly ask the children to connect the dots serially from a to z in figure 4(c) to complete the picture. This activity can be given for a period of 6 to 8 weeks for practice.

**Activity 4. Drawing**

Initially pencil design maker and stencils can be given for practice. This activity gives two hand coordination. The left hand can be used to hold the stencil firmly and right hand to move the pencil firmly. The computer aided drawing software helps in teaching step by step drawing process to complete the picture. The drawing tools in MS-word or paint brush software can also be used to practice drawing. This activity can be given for a period of 3 to 4 weeks.

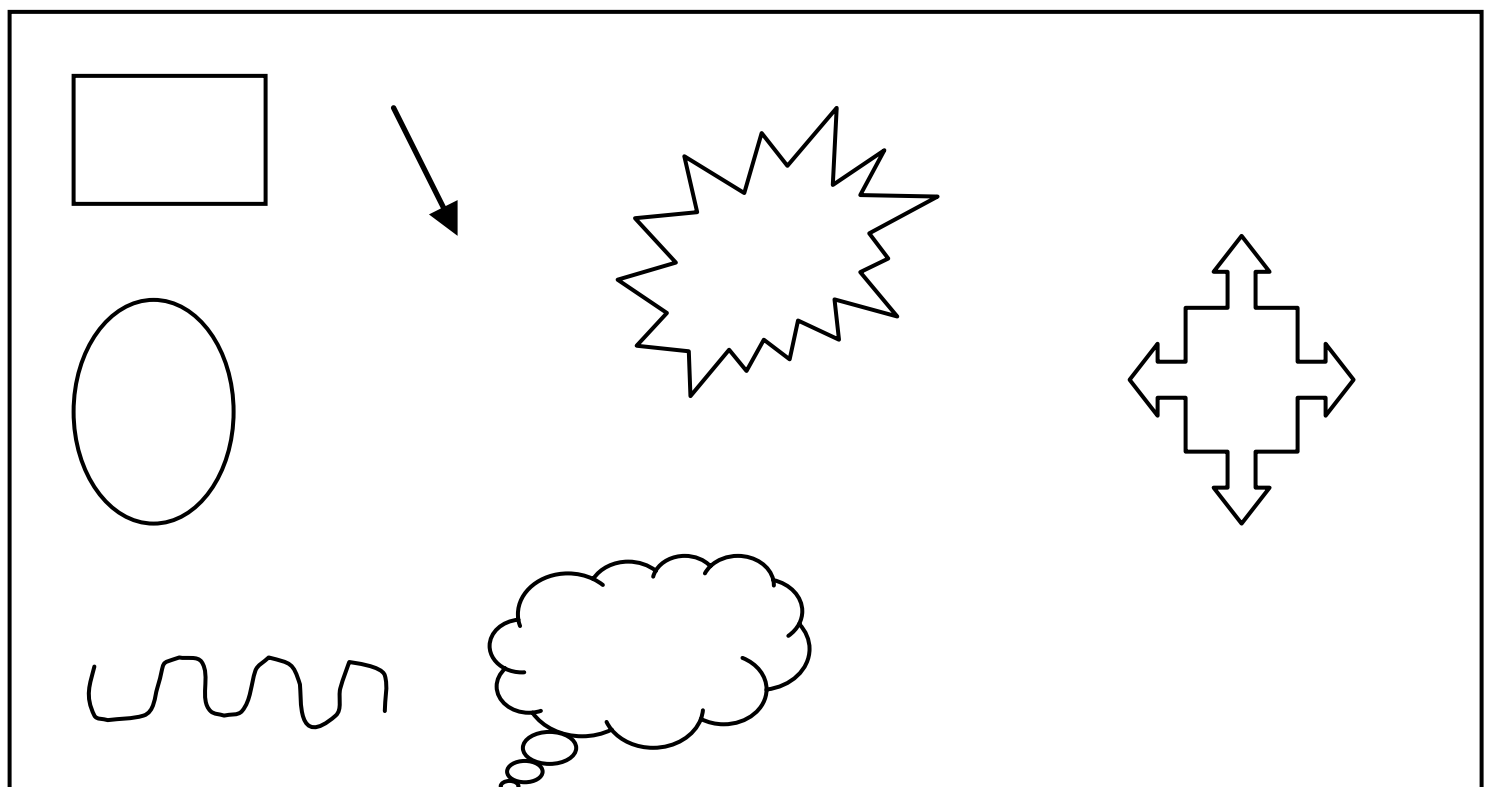

Figure 5. Pictures drawn using Drawing tools in MS-Word

**Activity 5. Games**

Several games specially designed for the kids can be given to improve the creativity, intelligence, reasoning and memory enhancement. **Memory card game** such as matching objects and shapes enhances the short term memory. In this game the cards are placed upside down on the screen. Make the child to click randomly on one card and ask to search for the matching pair. If the next card, randomly clicked, is not matching, try with another. The process is to be repeated until the match is found. Repeatedly remind the child to remember what he has seen and ask him/her to recall before clicking the next card. The game can also be played with pictures, shapes, objects or phonic words or match picture with word. Next, the **Maze games** where the children are supposed to move the cursor in the path specified by the hint and reach the destination through zigzag path. **Puzzle game** where a picture is shown on one side, the alphabets are displayed in random order on the other side. Ask the child to click the appropriate alphabet which is the starting letter of the picture. For example, if the picture of a cat is shown then the child has to click on C. Even a line can be drawn from the picture to matching alphabet. This game improves the matching concept. A group of pictures can be shown to ask the child to identify the **odd man out**. A pair of picture with slight difference can be shown and ask the child to **spot the differences**. This game increases the skills of observation and concentration. The pictures can be shown and a question can be asked to say yes / no. For example, by showing the picture of a cat, make the child answer yes/no by questioning whether it is a dog. The reasoning skill can be substantially developed. Show the objects related to opposite concept like **tall x short**, **big x small** etc., with pictures and ask the child to find out the opposite of each picture. This activity can be given for a period of 4-8 weeks for practice.

## Experimental Setup

During training period, weekly report has to be made separately for each child as an activity chart similar to as shown in Table 1 to record the performance of each child in each activity. The grading (1/2/3/4) can be used to know whether the child has done the activity independently or with partial / full assistance or refused to do the activity.

**Table 1. Activity Chart - I**

**Name of the Child:** **Age:**
**Date of commencement of the training:**

| Activity | Week1 | Week2 | Week3 | Week4 | Average Performance After a month |
|---|---|---|---|---|---|
| Coloring | 3 | 4 | 4 | 4 | |
| Placing or Pasting | 3 | 2 | 1 | 1 | |
| Tracing | 3 | 2 | 2 | 1 | |
| Drawing | | | | | |
| Creativity, reasoning, memory (Games) | | | | | |

4 – Independent 3 – with partial assistance
2 – with full assistance 1 – refused to do the activity

Compute,

**average performance after a month = (week1 + week2 + week3 + week4)/4**

Round off the average value. Similarly calculate the average value for each month on a separate sheet for a period of 6 to 12 months. Again take the average value of these values i.e., avg1, avg2, … avg12 as shown in table 2 and finally calculate the performance average as follows

**Performance average = ((avg1,avg2, ….. avg12)/12 ) * 100**

**Table 2. Activity Chart - II**

**Name of the Child:** **Age:**
**Date of commencement of the training:**

| Activity | Avg1 Month1 | Avg2 Month2 | Avg3 Month3 | Avg4 Month4 | Avg5 Month5 | Avg6 Month6 | Performance – average |
|---|---|---|---|---|---|---|---|
| Coloring | 3 | 4 | 4 | 4 | 3 | 4 | |
| Placing and Pasting | 2 | 2 | 1 | 1 | 3 | 2 | |
| Tracing | 3 | 2 | 1 | 1 | | | |
| Drawing | | | | | | | |
| Creativity, reasoning, memory (Games) | | | | | | | |

4 – Independent 3 – with partial assistance
2 – with full assistance 1 – refused to do the activity

Maintain separate activity sheet for each child. Considering the learning strategy in percentage before and after the commencement of the training a performance chart can be represented as shown in figure 6.

**Table 3. Learning Strategies in percentage**

| Activities | % of performance | |
|---|---|---|
| | **Before the Training** | **After the Training** |
| Coloring | 70 | 80 |
| Placing and Pasting | 60 | 70 |
| Tracing | 55 | 75 |
| Drawing | 60 | 70 |
| Creativity, reasoning, memory (Games) | 60 | 85 |

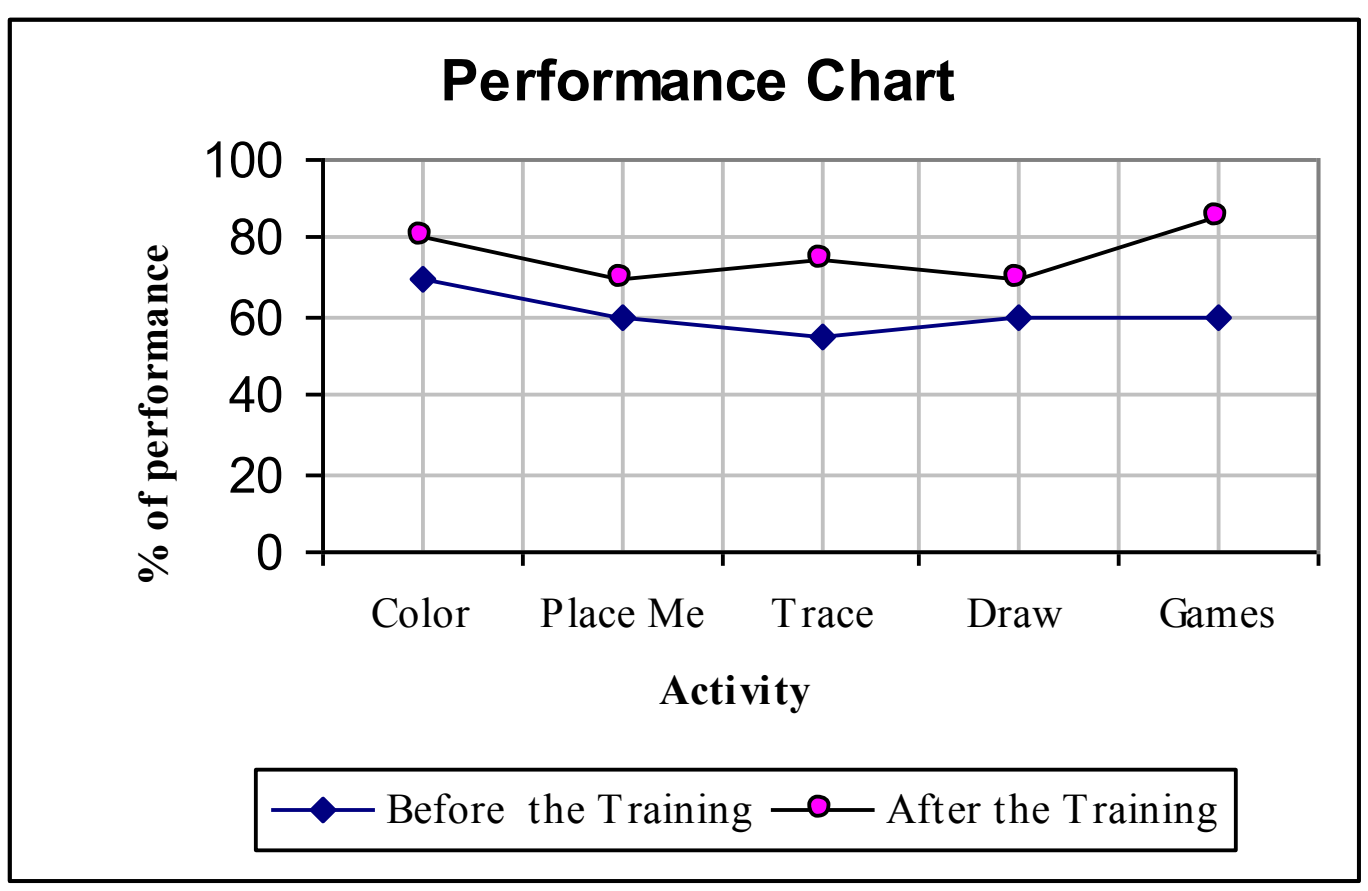

Figure 6. Performance Chart showing the Learning Strategy

After the training is completed, the feedback results as shown in Table 4 are taken from Parent, Special educator and Trainer. The overall percentage is calculated taking the average value of parent (p), special educator (s) and trainer (t). The results show how the technology supported in teaching and showed improvement in the learning process.

**Overall percentage = (p + s + t)/3**

**Table 4. Feedback results**

| The technology: | Parent p | Special Educator s | Trainer t | Overall Percentage |
|---|---|---|---|---|
| improved accuracy/quality | 3 | 3 | 3 | |
| increased speed/efficiency | 2 | 4 | 4 | |
| compensated for his/ her difficulty | 4 | 2 | 4 | |
| tapped into his/ her strengths | 3 | 4 | 3 | |
| was easy for /him/ her to learn | 4 | 2 | 2 | |
| was easy for him/ her to use | 2 | 1 | 2 | |

4 - Excellent (80-90%) 3 - Very good (70-80%)
2 - Good (60-70%) 1 - OK (50-60%)

## Conclusion and Future Work

In this paper we have presented various learning methodologies for wards with special needs using IT tools. Though our model designed and implemented requires internet and individual monitoring, it has shown better results or improvements in the learning process and the overall development of fine motor skills of the wards with special needs.

In future, we would be designing the models which will improve reading, writing, learning skills and also the concepts of basic mathematics, development of speech and language.

***Acknowledgement***

We would like to acknowledge website designers who have supported us in making the lesson plans with free downloadable educational softwares for kids and printable worksheets in pdf form. Sincere thanks to Dr. Varuni, leading psychologist, special educators, Speech and Hearing Centre, Manasagangotri, Mysore and NIMHANS, Bangalore for the guidance and valuable suggestions.